\documentclass[]{interact}

\usepackage{epstopdf}
\usepackage[caption=false]{subfig}

\usepackage[numbers,sort&compress]{natbib}
\bibpunct[, ]{[}{]}{,}{n}{,}{,}
\renewcommand\bibfont{\fontsize{10}{12}\selectfont}
\makeatletter
\def\NAT@def@citea{\def\@citea{\NAT@separator}}
\makeatother

\theoremstyle{plain}

\theoremstyle{definition}

\theoremstyle{remark}

\begin{document}

\articletype{RESEARCH ARTICLE}

\title{Fertile metastability}

\author{
\name{P. Pieranski\textsuperscript{a}\thanks{CONTACT P. Pieranski. Email: pawel.pieranski@u-psud.fr} and M.H. Godinho\textsuperscript{b}}
\affil{\textsuperscript{a}Laboratoire de Physique des Solides, Université Paris-Saclay, Orsay, France; \textsuperscript{b}i3N/CENIMAT, NOVA University of Lisbon, Portugal}
}

\maketitle

\begin{abstract}
We deal here with three metastable systems: 1$^{\circ}$- the dowser texture, 2$^{\circ}$- supertwisted cholesterics and 3$^{\circ}$- hypotwisted cholesterics. We outline remarkable properties of tropisms of the dowser texture stemming from its low symmetry and we show that, using setups called Dowsons Colliders, nematic monopoles can be nucleated, set into motion and brought into collisions in the dowser texture. Subsequently we point out that nucleation of dislocation loops occurs in cholesteric layers compressed or dilated between cylindrical mica sheets. Under compression, three modes of nucleation of dislocation loops have been identified: \emph{individual}, \emph{serial} and \emph{continuous}. The serial mode generates dislocation nets made of L circular loops connected by C radial crossing into superloops. Similar superloops can also be generated under dilation. We analyse the topology of superloops in terms of the theory of knots and point out they can be reduced into the unknot by Reidemeister moves. In other words, superloops are multiply folded loops that can be continuously unfolded.
\end{abstract}

\begin{keywords}
metastability, dowser texture, dislocations, cholesterics, knots
\end{keywords}

\section{Introduction}
\subsection{Considerations about metastability inspired by a view from the Caparica shore}
\begin{figure*}
\begin{center}
\includegraphics[width=4.5 in]{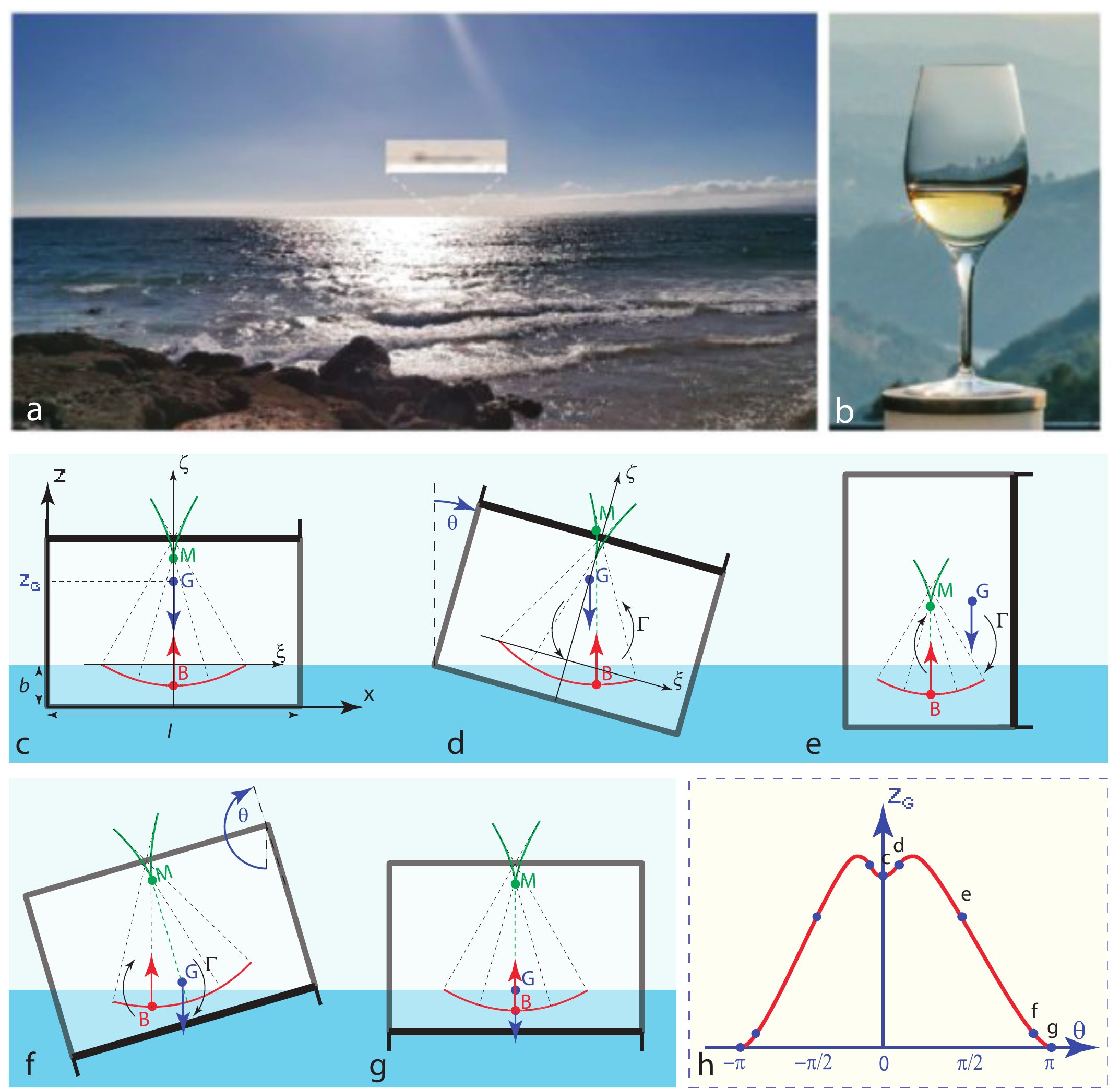}
\caption{Metastability. a) A view of the ocean from the shore at Caparica. b) Sparkling vinho verde. c) Metastable position of a ship with a rectangular cross section. G is the center of gravity of the ship, B - the center of buoyancy, M - the metacenter. d) Negative restoring torque corresponding to a small enough roll angle $\theta$. d) Positive destabilising torque corresponding to large enough roll angles $\theta$. g) Stable position of the ship after capsizing. h) Potential energy of the ship versus the roll angle $\theta$.       }
\label{fig:metastability_ship}
\end{center}
\end{figure*}

The ILCC2022 held in the NOVA School of Science and Technology located in Caparica, a suburb of Lisbon. The splendid view on the Atlantic ocean from the Caparica shore shown in Fig.\ref{fig:metastability_ship}a accompanied by the slightly sparkling taste of vinho verde ( Fig.\ref{fig:metastability_ship}b), led participants of the conference to deviate from the topical discussion about metastable liquid crystal systems to more general considerations about the concept of metastability in physics.

The thin band of white clouds floating above the horizon line in Fig.\ref{fig:metastability_ship}a is obviously related to nucleation of water droplets in supersaturated water vapors similar to nucleation of gas bubbles in a supersaturated solution of carbon dioxide in vinho verde. These two canonical examples of the thermodynamic metastability are well known.

Another interesting eye-catching detail - the presence of a huge cargo ship visible in Fig.\ref{fig:metastability_ship}a at a higher magnification - is related to stability of floating bodies. Let us consider the example of a ship with rectangular cross section represented schematically in Fig.\ref{fig:metastability_ship}c. As its center of gravity G is located about the center of buoyancy B, one could think naively that this position of the ship is unstable with respect to roll. However, upon a small enough roll angle $\theta$, the buoyancy center moves on the buoyancy curve (drawn in red) to the right so much that the buoyancy and gravity forces generate a negative torque $\Gamma$ tending to reduce the roll of the ship (see Fig.\ref{fig:metastability_ship}d). However, when the roll angle $\theta$ is too large, the sign of the torque $\Gamma$  becomes positive so that the roll angle grows and the ship capsizes. In the upside-down position (Fig.\ref{fig:metastability_ship}e), the potential energy of ship is lower than in the initial position (Fig.\ref{fig:metastability_ship}e) that therefore must be considered as metastable with respect to capsizing.

The buoyancy curve is locus of buoyancy centers B, that is to say, centers of mass of water displaced by the ship. In  Fig.\ref{fig:metastability_ship}c (the red curve), it has been calculated using expressions taken from the paper of M\'{e}gel and Kliava \cite{Kliava_ship} :
\begin{eqnarray} \label{eq:bouyancy}
  x_{B}(\theta) &=& \frac{1}{12}\frac{l^{2}}{b}tan\theta \\
  z_{B}(\theta) &=& -\frac{1}{2}b+\frac{1}{24}\frac{l^{2}}{b}tan^{2}\theta
\end{eqnarray}
with $l=6$ and $b=1$. In the same Fig.\ref{fig:metastability_ship}c, the green curve called metacentric curve M is evolute of the buoyancy curve i.e. locus of centers of curvature of B. Invented in 18$^{th}$ century by Pierre Bouguer \cite{Bouguer}, it is helpful in determination of the stability of ships: as long as the gravity center G of the ship is located below the metacenter M, the ship will not capsize.
In  Fig.\ref{fig:metastability_ship}c, the green metacentric curve was calculated using expressions taken from the paper of M\'{e}gel and Kliava \cite{Kliava_ship} :
\begin{eqnarray} \label{eq:bouyancy}
  x_{M}(\theta) &=& -\frac{1}{12}\frac{l^{2}}{b}tan^{3}\theta \\
  z_{M}(\theta) &=& -\frac{1}{2}b+\frac{1}{24}\frac{l^{2}}{b}\left(\frac{3}{cos^{2}\theta}-1\right)
\end{eqnarray}
with $l=6$ and $b=1$.

Even if it is potentially dangerous, the metastability of ships can be tamed and made fertile: nowadays, huge and tall container ships transport all kind of fret all around the world.

Let us emphasize that if the gravity center G in Fig.\ref{fig:metastability_ship}c was located above the metacenter M, this position of the ship would correspond to the maximum of the gravitational potential and therefore would be unstable - the ship would inevitably capsize.
\subsection{Stability of twist and splay-bend distortions}

Let us examine now stability of the twist and splay-bend distortions in a nematic slab of thickness h with planar and homeotropic boundary conditions represented in Fig.\ref{fig:metastability_distorsions}.
\subsubsection{Twist distortion}
\begin{figure*}
\begin{center}
\includegraphics[width=5 in]{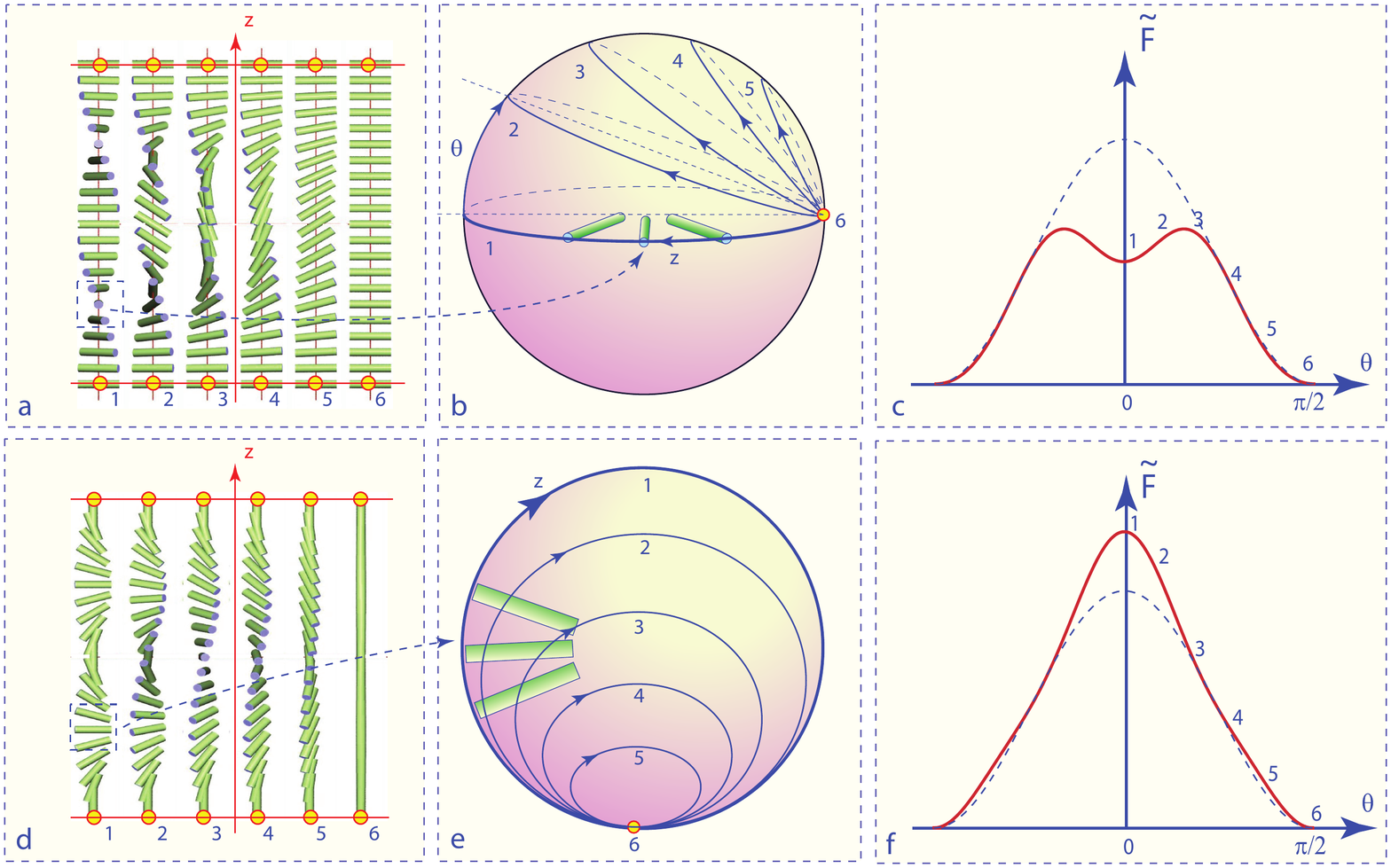}
\caption{Stability of the splay-bend and twist distortions. a) Continuous suppression of the twist distortion. b) Mapping of the six stages of the distortion represented in (a) onto the North hemisphere - the space of the nematic order parameter. c) Variation of the distortion energy density per unit area $\tilde{F}$ in the approximation of isotropic elasticity (dashed blue line) and in the anisotropic case of 5CB where $K_{22}/K_{11}\approx 0.5$ and $K_{33}/K_{11}\approx 1.5$ (solid red line). d) Continuous suppression of the splay-bend distortion. e) Mapping of distortions represented in (d) onto the East hemisphere - the space of the nematic order parameter. c) Variation of the distortion energy density per unit area calculated in the isotropic approximation (dashed blue line) and in the case of 5CB with $K_{22}/K_{11}\approx0.5$ and $K_{33}/K_{11}\approx1.5$ (solid red line).}
\label{fig:metastability_distorsions}
\end{center}
\end{figure*}

In the case of the twist distortion in Fig.\ref{fig:metastability_distorsions}a1, the director rotates around the z axis by $2\pi$ between the limit surfaces equipped with the planar anchoring. This director field $\mathbf{n}(z)$ for z between 0 and h,  maps onto the equator of the nematic order parameter space (North hemisphere) shown in Fig.\ref{fig:metastability_distorsions}b. Orientations $\mathbf{n}(0)$ and $\mathbf{n}(h)$ of the director on limit surfaces map onto the same point (1,0,0) represented by the yellow circle.

It is well known that for topological reasons such a distortion can be suppressed by a continuous deformation of the director field represented in real space in the series of six drawings 1-6 in Fig.\ref{fig:metastability_distorsions}a. Mapped onto the unit sphere such a deformation corresponds to a series of shrinking circles passing through the point (1,0,0) and tilted with respect to the equator by the angle $\theta$ growing from $0$ to $\pi/2$. Analytically the evolving director field can be expressed as
\begin{equation} \label{eq:twist}
\mathbf{n}(z,\theta)=\left[cos^{2}(\theta)cos(2\pi  z)-sin^{2}(\theta),cos(\theta) sin(2\pi  z),cos(\theta)sin(\theta)(1+cos(2 \pi z))\right]
\end{equation}

Using this expression of the director field $\mathbf{n}(z,\theta)$, one can calculate the density per unit area of the distortion energy:
\begin{equation} \label{eq:energy_twist}
F=\frac{K_{11}}{2}\int_{0}^{h}\left((\mathbf{div}\mathbf{n})^{2}+\frac{K_{22}}{K_{11}}(\mathbf{n}\cdot \mathbf{rot}\mathbf{n})^{2}+\frac{K_{33}}{K_{11}}(\mathbf{n}\times \mathbf{rot}\mathbf{n})^{2}\right)dz
\end{equation}
In the approximation of the isotropic elasticity i.e. for $\frac{K_{22}}{K_{11}}=\frac{K_{33}}{K_{11}}=1$, one obtains
\begin{equation} \label{eq:energy_twist_isotropic}
\tilde{F}=\frac{F}{4\pi^2 h K_{11}/2}=cos^2(\theta)
\end{equation}
which means that the pure twist distortion for $\theta=0$ corresponds to the maximum of the energy density $F(\theta)$  (see the dashed blue line in Fig.\ref{fig:metastability_distorsions}c) and therefore is unstable with respect to the continuous transformation involving the splay and bend distortions.

In the general case of an anisotropic elasticity, the expression of the energy density becomes :
\begin{equation} \label{eq:energy_twist_anisotropic}
\tilde{F}=2 \pi^2 cos^2(\theta) \left(\frac{K_{22}}{K_{11}} (2 cos^{4}(\theta)+sin^{4}(\theta)) +3 \frac{K_{33}}{K_{11}} cos^2(\theta) sin^2(\theta)+sin^2(\theta)  \right)
\end{equation}
In the case of 5CB, where $\frac{K_{22}}{K_{11}}\approx 0.5$ and $\frac{K_{33}}{K_{11}}\approx 1.5$, the function $\tilde{F}(\theta)$ (see the red solid line in Fig.\ref{fig:metastability_distorsions}c) has a local minimum at $\theta=0$ separated by an energy barrier from the absolute minimum at $\theta=\pi/2$. Therefore the pure twist distortion in 5CB is not absolutely unstable but only metastable. This last conclusion remains valid when the angle of rotation of the director between $z=0$ and $z=h$ is an integer multiple of $2\pi$.

In section \ref{sec:superloops} we will point out that the metastability of the twist distortion in supertwisted cholesterics leads to generation of loops of the so-called thick dislocation lines.
\subsubsection{Splay-bend distortion}
In the case of the splay-bend distortion in Fig.\ref{fig:metastability_distorsions}d1, the director rotates around the y axis by $2\pi$ between the limit surfaces equipped with the homeotropic anchoring. This director field $\mathbf{n}(z)$ for z between 0 and h,  maps onto the 0 and 180$^{\circ}$ meridians of the nematic order parameter space (East hemisphere) shown in Fig.\ref{fig:metastability_distorsions}e. Orientations $\mathbf{n}(0)$ and $\mathbf{n}(h)$ of the director on limit surfaces map onto the same point (0,0,-1) represented by the yellow circle.

For the same topological reasons as above, such a distortion can be suppressed by a continuous deformation of the director field represented in real space in the series of six drawings 1-6 in Fig.\ref{fig:metastability_distorsions}d. Mapped onto the unit sphere (see Fig.\ref{fig:metastability_distorsions}e) such a deformation corresponds to a series of shrinking circles passing through the point (0,0,-1) and tilted with respect to the z axis by the angle $\theta$ growing from $0$ to $\pi/2$. Analytically the evolving director field can be expressed as
\begin{equation} \label{eq:splay_bend}
\mathbf{n}(z,\theta)=\left[cos(\theta)cos(2\pi  z),cos(\theta)sin(\theta)(1- sin(2\pi  z)),sin^{2}(\theta)+cos^{2}(\theta)sin(2\pi  z)\right]
\end{equation}
Using this expression of the director field we calculated variation of the elastic energy density $\tilde{F}$ with the angle $\theta$. Results obtained in the isotropic approximation and in the case of 5CB are plotted respectively with the blue dashed line and with the solid red line in Fig.\ref{fig:metastability_distorsions}f. As expected, contrary to the case of the twist distortion analysed above, the elastic anisotropy of 5CB does not produce the local minimum at $\theta=0$ but makes the maximum higher.

In conclusion, in 5CB, the $2\pi$ splay-bend distortion lodged between surfaces with homeotropic anchoring is absolutely unstable. However, if the angle of rotation is not $2\pi$ but $\pi$ (see Fig.\ref{fig:dowser}a), the director field $\mathbf{n}(z)$ maps on just one meridian of the order parameter space so that the splay-bend distortion cannot be suppressed otherwise as by motion of disclinations transforming the splay-bend distortion into the homogeneous homeotropic texture.

\section{Remarkable properties of the dowser texture}
\label{sec:dowser}
It is well known since decades that nematic samples prepared between parallel glass slides equipped with homeotropic anchorings contain, immediately after filling, the $\pi$ splay-bend and homeotropic textures separated by disclination loops that were believed to expand always with the benefit of the homeotropic texture. For this reason the $\pi$ splay-bend texture was considered as unstable and little attention was paid to it. However, if the homeotropic domains were missing, the $\pi$ splay-bend texture would be not unstable but only metastable with respect to nucleation of homeotropic domains.

As explained in the chapter 4 "Physics of the dowser texture" in the collective book "Liquid Crystals - New perspectives" published recently, nucleation of the homeotropic domains is hindered by the energy barrier so efficiently that in certain conditions the $\pi$ splay-bend distortion can be preserved indefinitely \cite{PP_MHG_English,PP_MHG_French}.
\subsection{Tropisms}
\label{sec:tropisms}
\begin{figure*}
\begin{center}
\includegraphics[width=5 in]{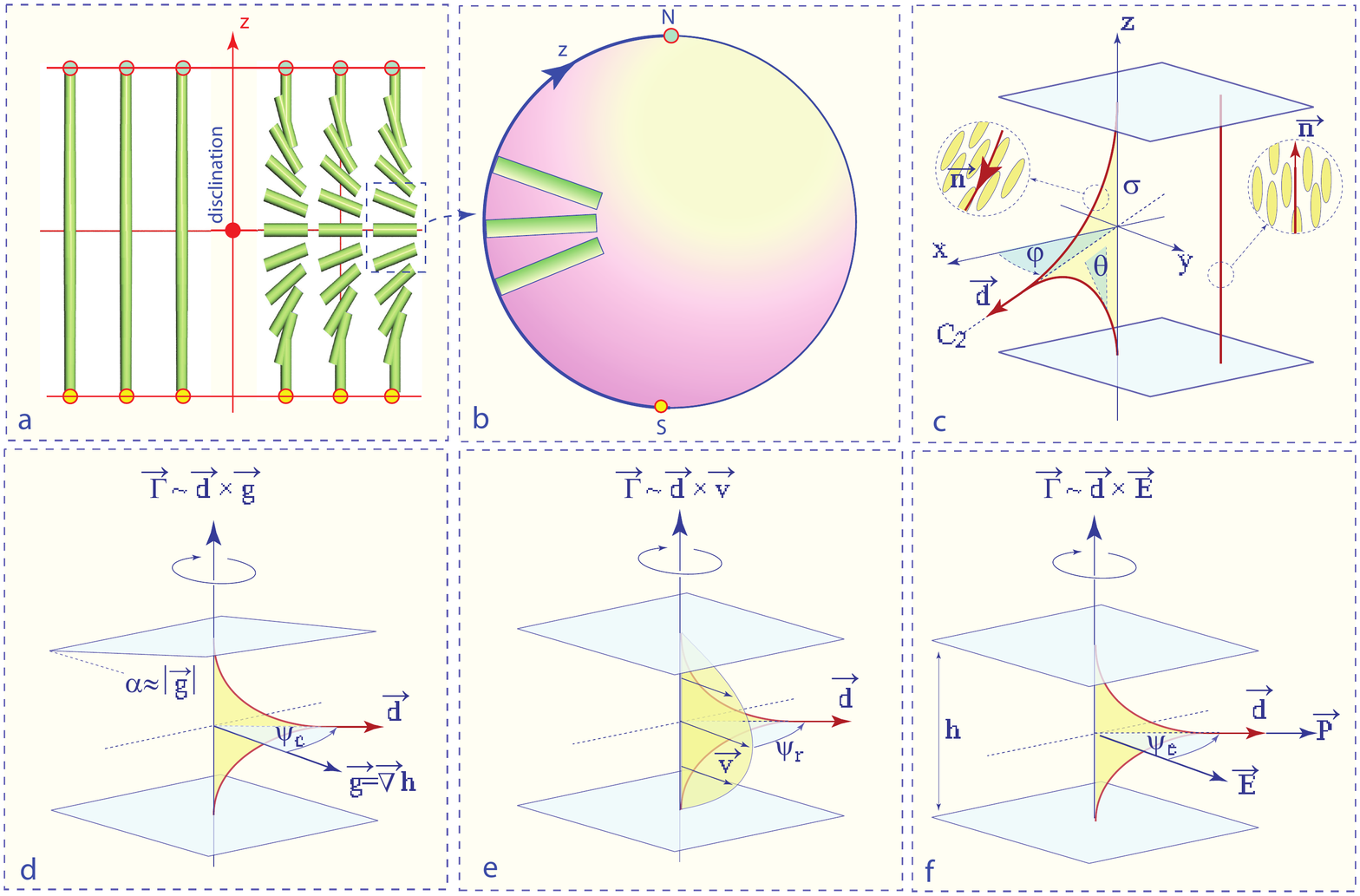}
\caption{Tropisms of the dowser texture. a) The director fields of the homeotropic and dowser textures separated by a disclination. b) Mapping of the dowser texture on one meridian in the space of the nematic order parameter. c) Perspective view of the homeotropic and dowser textures. Symmetries of the dowser texture: mirror plane $\sigma$ and the twofold axis $C_{2}$. Order parameter of the dowser texture: the unitary 2D vector $\vec{d}$. When the limit plates are parallel, the azimuthal angle $\varphi$ is arbitrary. d) Cuneitropism. e) Rheotropism. f) Electrotropism.}
\label{fig:dowser}
\end{center}
\end{figure*}
\begin{figure}
\begin{center}
\includegraphics[width=5 in]{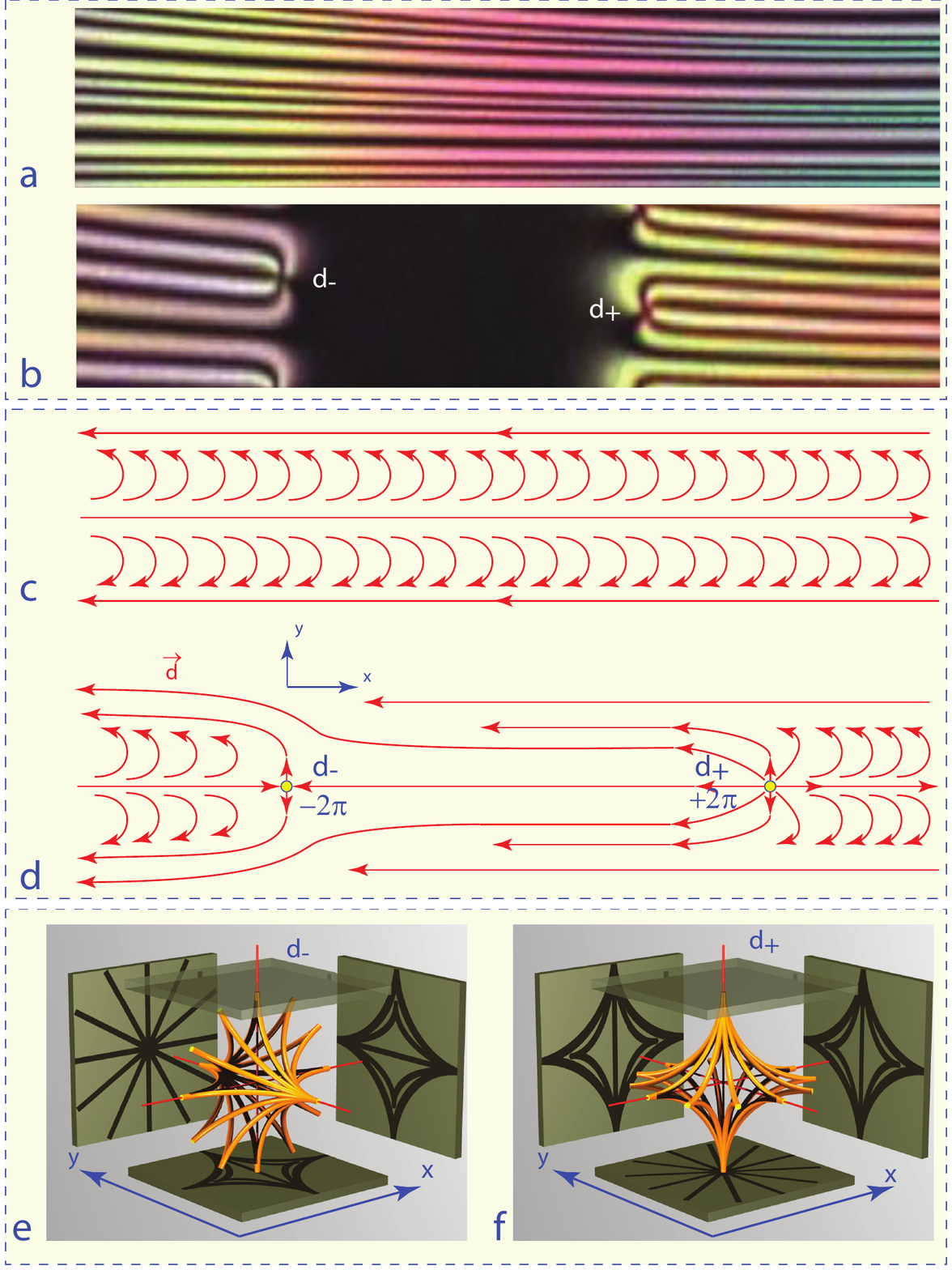}
\caption{Dowsons d+ and d-, the $+2\pi$ and $-2\pi$ point defects of the dowser field $\vec{d}$, as monopoles of the director field in a 3D nematic. a) View in the polarising microscope of a wound up dowser field that can be seen as a stack $2\pi$ walls. b) Generation of pairs of dowsons d+ and d- by breaking of the $2\pi$ walls. c) Dowser field of a $2\pi$ wall in the wound up dowser texture. d) Dowser field in the  presence of the (d+d-) pair. e-f) View of the 3D director fields in vicinity of nematic monopoles corresponding to dowsons d+ and d-. Let us stress that topologically the two fields are identical but they differ by their orientations with respect to the limit surfaces.          }
\label{fig:dowsons}
\end{center}
\end{figure}
Such a long lived $\pi$ splay-bend distortion was dubbed "the dowser texture" because of the resemblance of its director field with the wooden tool of dowsers. It has been pointed out that, due to its low symmetry $C_{2v}$  and degeneracy with respect to rotations around the z axis (see Fig.\ref{fig:dowser}c), the dowser texture is endowed with tropisms i.e. first order sensitivities to vector fields such thickness gradients (cuneitropism, see Fig.\ref{fig:dowser}d), Poiseuille flows  (rheotropism, see Fig.\ref{fig:dowser}e) and electric fields (electrotropism, see Fig.\ref{fig:dowser}f) (see references \cite{PP_MHG_English} and \cite{PP_MHG_French}).
\subsection{Natural universe of nematic monopoles}
\label{sec:monopoles}
Beside its tropisms, the dowser texture has another remarkable property: it can be considered as a natural universe of nematic monopoles \cite{Kleman_Lavrentovich_2006} that can be generated, set into motion and brought to collisions by means of the so-called dowsons colliders. In the example shown in Fig.\ref{fig:dowsons} the dowser fields is first wound up (see Figs.\ref{fig:dowsons}a and c) which means that its phase varies with the y coordinate. Let us note that such a wound up dowser field can be seen as a system of $2\pi$ walls.

A rapid enough transitory Poiseuille flow applied in the y direction breaks the $2\pi$ walls (see Figs.\ref{fig:dowsons}b and d) and consequently generates pairs of $+2\pi$ and $-2\pi$ point defects of the 2D dowser field $\vec{d}$. In 3D, the director field of these defects, dubbed dowsons d+ and d-, has structure of nematic monopoles depicted in Figs.\ref{fig:dowsons}e and f. Let us stress that structures of monopoles corresponding to both dowsons d+ and d- are here identical (hyperbolic) but their orientations are different: in the dowsons d+ and d-, the symmetry axis of the hyperbolic monopole is respectively parallel to the z and y axes .

For a much more detailed and exhaustive discussion of properties of the dowser texture we recommend the chapter 4 in the recently published collective books \cite{PP_MHG_English,PP_MHG_French}.

\section{Nucleation of dislocation loops in supertwisted cholesterics}
\label{sec:nucleation_supertwisted}
\subsection{Crossed cylinders geometry}
\label{sec:crossed_cylinders}
Since their discovery in 1921 by F. Grandjean \cite{Grandjean_1921}, dislocations in Grandjean-Cano wedges have been abundantly studied in past. Many articles as well as book chapters have been devoted to them. In particular, contributions of Maurice Kleman in this field were numerous and crucial \cite{Kleman_Friedel_69,Kleman_PLP,Kleman_Progress_89,Friedel_Kleman,Kleman_Lavrentovich,Toulouse_Kleman}.

\begin{figure*}
\begin{center}
\includegraphics[width=4 in]{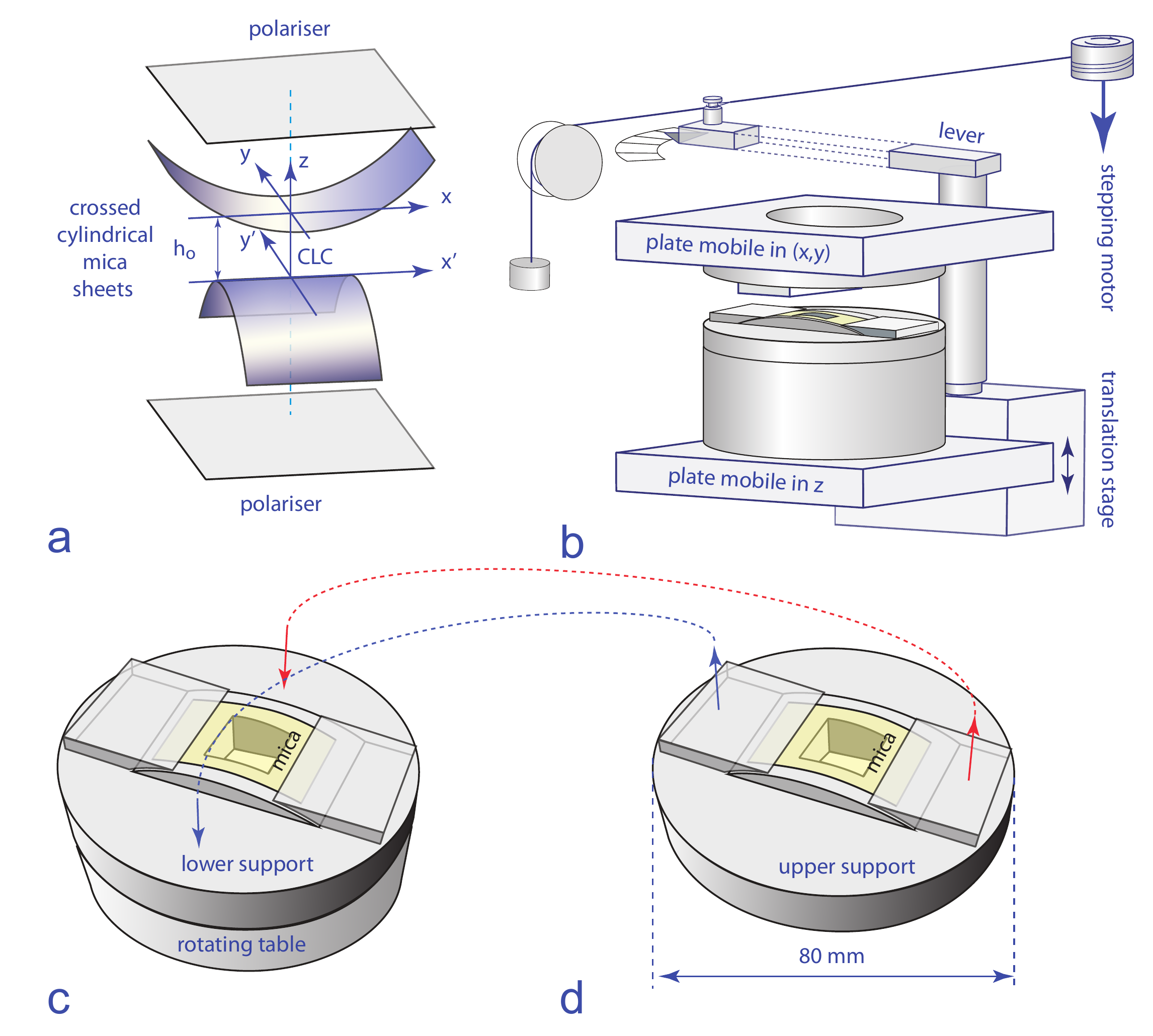}
\caption{Setup for nucleation and observation of dislocations in cholesterics. a) Geometry of the cylinder/cylinder gap. b) Perspective view of the setup. c and d) Lower and upper supports of the mica sheets. }
\label{fig:setup}
\end{center}
\end{figure*}

\begin{figure*}
\begin{center}
\includegraphics[width=5.5 in]{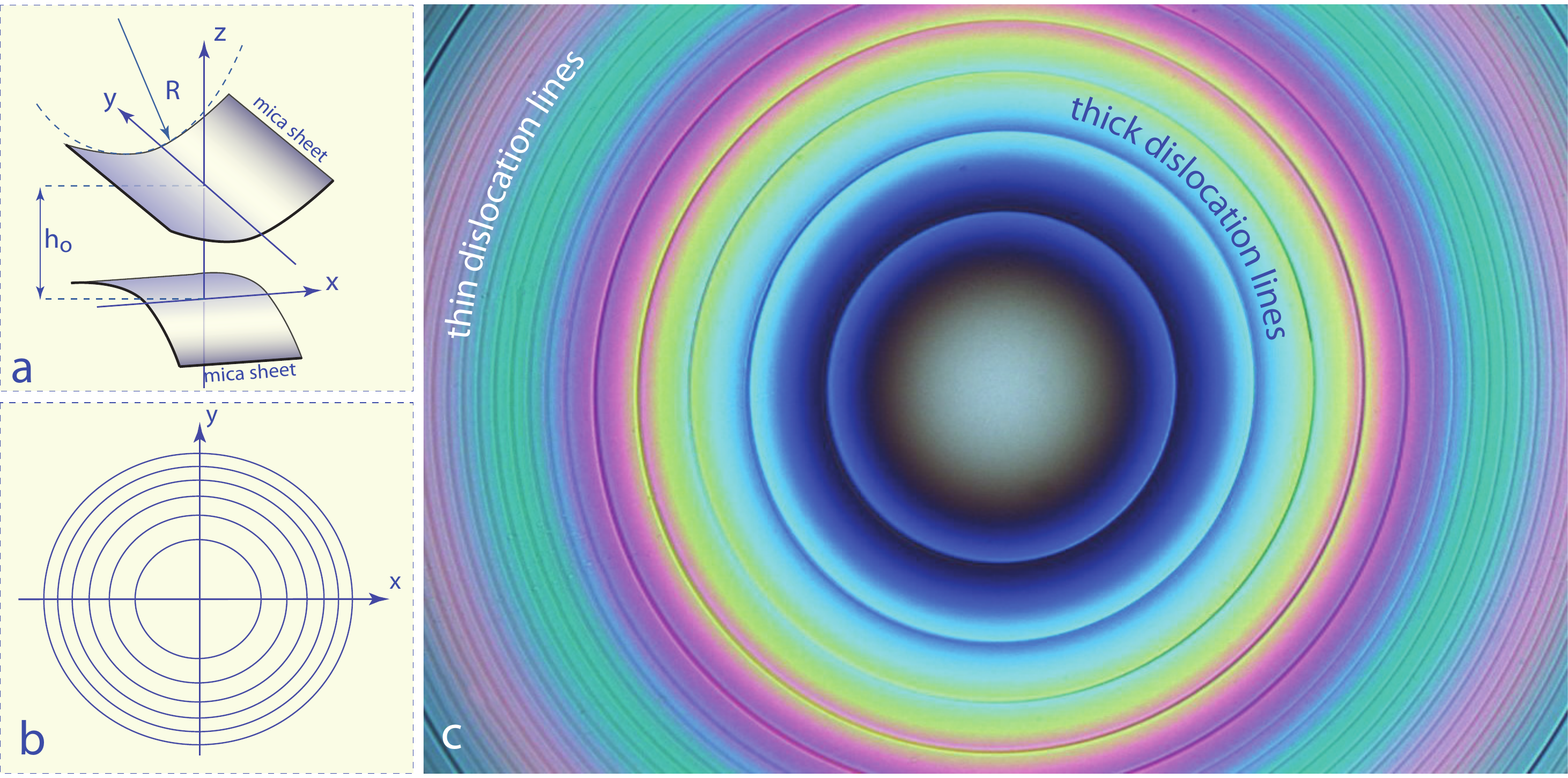}
\caption{Dislocation lines in the gap between two crossed cylindrical surfaces made of cleaved mica sheets. a) Perspective view of the wedge. b) Top view of circular level lines in the wedge. c) Unexpected system of dislocation lines: contrary to common beliefs \emph{thin} lines are located in the \textbf{thick} periphery of the wedge while \textbf{thick} lines appear in the \emph{thin} central part of the wedge.  }
\label{fig:deux_cylindres}
\end{center}
\end{figure*}

With the aim to illustrate the article \cite{Pieranski_Memorial} published in the Memorial Issue of Liquid Crystals Reviews in honor of Maurice Kleman with pictures of dislocations in colour, we performed recently new experiments with a setup in which the wedge is formed by two crossed cylindrical mica sheets (see Fig.\ref{fig:setup}). In this geometry, dislocation lines have shapes of concentric circular loops (see Fig.\ref{fig:deux_cylindres}c) parallel to level lines (see Fig.\ref{fig:deux_cylindres}b) in agreement with recent studies of Zappone and Bartolino \cite{Zappone_Bartolino}. Our setup, tailored for the purpose of optical observations, allowed us to unveil new unexpected facts concerning nucleation of dislocation loops.
\begin{figure*}
\begin{center}
\includegraphics[width=5.5 in]{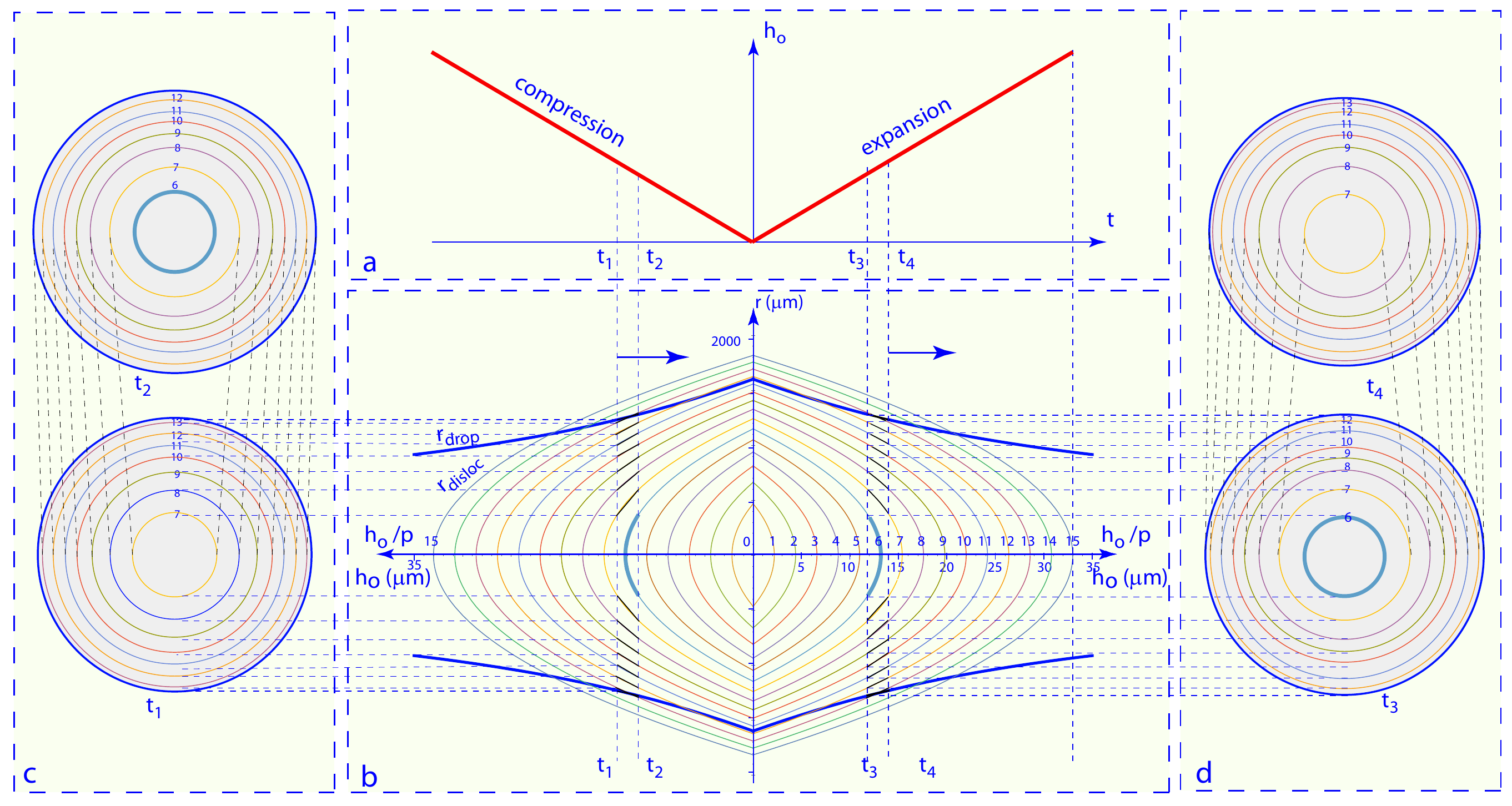}
\caption{Hypothetical evolution of dislocation loops in the absence of nucleation barriers. a) Variation of the minimal thickness $h_{o}$ in time. b) Variation of the radii of dislocations $r_{disloc}$ and of the radius of the droplet $r_{drop}$ with $h_{o}$ given respectively by equations \ref{eq:rdisloc} and \ref{eq:rdrop}. c) Evolution of dislocation loops in the time interval $(t_{1},t_{2})$ during compression of the gap between cylindrical mica sheets. Note nucleation of the loop n=6. d) Evolution of dislocation loops in the time interval $(t_{3},t_{4})$ during expansion of the gap. Note collapse of the loop n=6.}
\label{fig:compression_expansion}
\end{center}
\end{figure*}
\subsection{Systems of individual dislocation loops}
\label{sec:crossed_cylinders}
The first one concerns the distribution of the so-called \emph{thin} and \textbf{\emph{thick}} lines corresponding to dislocations with Burgers vectors respectively equal to the half pitch p/2 and the whole cholesteric pitch p. In the example shown in Fig.\ref{fig:deux_cylindres}c, \emph{thin} and \textbf{\emph{thick}} lines are located respectively in \textbf{\emph{thick}} and \emph{thin} parts of the wedge in contradiction with expectations resulting from calculations of the elastic energy (see for example in ref. \cite {Oswald_Pieranski}). Indeed, the absolute minimum of the elastic energy of the system of circular dislocations would be reached if the \emph{thin} and \textbf{\emph{thick}} lines were located respectively in \emph{thin} and \textbf{\emph{thick}} parts of the crossed cylinders wedge.

Therefore, the system of dislocations shown in Fig.\ref{fig:deux_cylindres}c is doubly metastable: the \textbf{\emph{thick}} lines should split into pairs of \emph{thin} lines and the pairs of \emph{thin} lines should merge into \textbf{\emph{thick}} lines. These splitting and merging processes are obviously hindered by prohibitive energy barriers.

All dislocation lines in Fig.\ref{fig:deux_cylindres}c form separated loops because they were nucleated one after another during compression of the wedge i.e. during reduction of the minimal distance $h_{o}$ between the mica sheets. The thin lines, located in the wedge periphery, were nucleated first on some imperfection of one of the mica surfaces. The thick lines were nucleated subsequently on another imperfection of mica surfaces. We postpone a more detailed discussion of the nucleation of individual loops to another paper.
\subsection{Serial nucleation of dislocation loops linked into superloops}
\label{sec:crossed_cylinders}
The second surprising result consists in the discovery of two other modes of nucleation of dislocation loops: the serial mode and the Frank-Read-like continuous mode \cite{Pieranski_Memorial}. Here we will deal only with the serial mode.
\begin{figure*}
\begin{center}
\includegraphics[width=5 in]{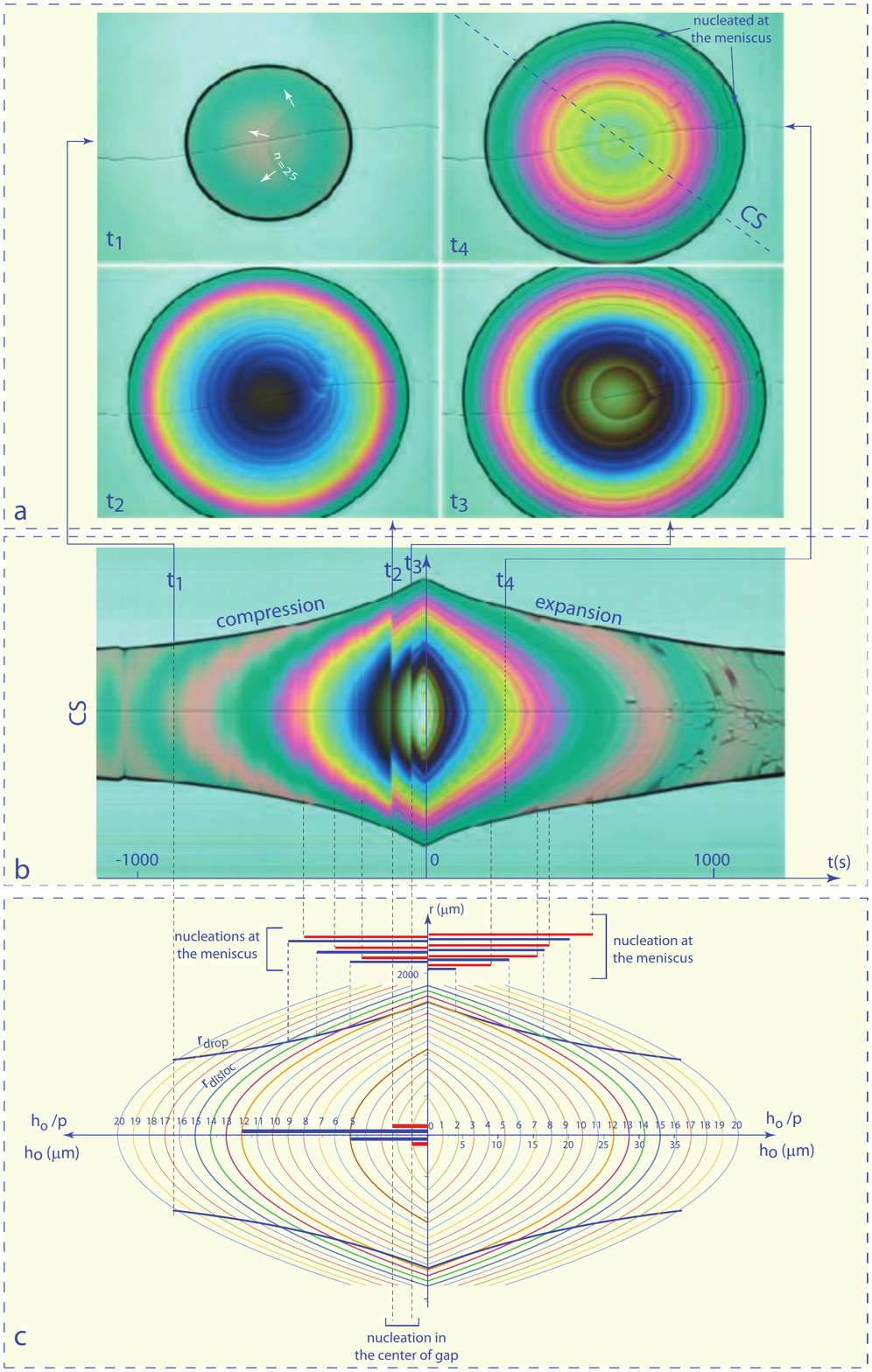}
\caption{Evolution of dislocation loops in a real experiment. a) Snapshots taken at instants $t_{1}-t_{4}$ defined in b. b) Spatio-temporal cross section extracted from a video along the line CS defined in the picture a-t4. c) Interpretation of events in the spatio-temporal cross section.}
\label{fig:serial}
\end{center}
\end{figure*}
The typical experiment in which the serial nucleation of dislocation loops occurs consists in a slow compression of the wedge i.e. reduction of its minimal thickness $h_{o}$ (see Figs.\ref{fig:compression_expansion} and \ref{fig:serial}).
\subsubsection{Evolution of dislocations in the absence of nucleation barriers}
\label{sec:no_barriers}
Before discussing results of such an experiment, let us imagine how dislocation lines would evolve in the absence of nucleation barriers (see Fig.\ref{fig:compression_expansion}). As the volume of the cholesteric droplet contained between crossed cylindrical mica sheets remains constant, its radius $r_{drop}$ grows accordingly to the expression:
\begin{equation}\label{eq:rdrop}
r_{drop}=p_{o}\sqrt{\sqrt{4 \tilde{h}_{o}^{2} \tilde{R}^{2} + \tilde{r}_{max}^{4}}-2 \tilde{h}_{o} \tilde{R}}
\end{equation}
resulting from geometrical consideration in which $\tilde{r}_{max}=r_{max}/p_{o}$ represents the maximal radius of the drop reached for $\tilde{h}_{o}=h_{o}/p_{o}=0$ (see the thick blue line labeled $r_{drop}$ in Fig.\ref{fig:compression_expansion}b) and $\tilde{R}=R/p_{o}$. Simultaneously, the radii of dislocation loops already present in the cylinder/cylinder wedge will grow accordingly to the expression
\begin{equation}\label{eq:rdisloc}
r_{disloc}(N,\tilde{h}_{o})=p_{o}\sqrt{(N-\tilde{h}_{o}) 2\tilde{R}}
\end{equation}
in which $N$ is the index of dislocations, $\tilde{h}_{o}$ and $\tilde{R}$ are respectively the minimal thickness and the radius of curvature of mica sheets expressed in units of the equilibrium cholesteric  pitch $p_{o}$. The set of thin coloured lines labeled from 1 to 15 in Fig.\ref{fig:compression_expansion}b was drawn using this expression. The line with the index N=15 is also labeled $r_{disloc}$.

The expression \ref{eq:rdisloc} is solution of the equation
\begin{equation}\label{eq:FPK1}
F_{P-K}=0
\end{equation}
concerning the Peach-Koehler force per unit length acting on dislocations. In our case of the cholesteric samples, the Peach-Koehler force is equal to the difference between the densities per unit area of the elastic distortion in fields separated by dislocations :
\begin{equation}\label{eq:FPK2}
F_{P-K}=[f_{el}(N+1)-f_{el}(N)]h
\end{equation}
where h is the local thickness and
\begin{equation}\label{eq:FPK3}
f_{el}(N)=\frac{K_{22}}{2}\left(\frac{2\pi}{p_{o}}\right)^2 \left( \frac{N-1/2}{\tilde{h}_{o}+\tilde{r}^2/2\tilde{R}}- 1\right)^2
\end{equation}
is the density per unit volume of the elastic distortion. From equations \ref{eq:FPK2} and \ref{eq:FPK3} one obtains
\begin{equation}\label{eq:FPK4}
F_{P-K}=\frac{2K_{22}\pi^2}{p_{o}}\left(\frac{N}{\tilde{h}_{o}+\tilde{r}^2/2\tilde{R}}-1\right)
\end{equation}
Finally, from the condition of equilibrium \ref{eq:FPK1} one gets the expression \ref{eq:rdisloc}.

The lower and upper schemes of  Fig.\ref{fig:compression_expansion}c represent the drop during the compression stage respectively at times t1 and t2 defined in Fig.\ref{fig:compression_expansion}b. Let us stress that in the absence of nucleation barriers the loop N=6 nucleates when the minimal gap thickness $h_{o}$ reaches the critical value given by $r_{disloc}(6,h_{o})=0$ (eq.\ref{eq:rdisloc}).

Similarly, the lower and upper schemes of  Fig.\ref{fig:compression_expansion}d represent the drop during the expansion stage respectively at times t3 and t4 defined in Fig.\ref{fig:compression_expansion}b. Note that the loop N=6 collapses in the time interval $(t_{3},t_{4})$.
\subsubsection{Evolution of the net of dislocations in real experiments}
\label{sec:with_barriers}
The most striking feature of Fig.\ref{fig:compression_expansion}b is its mirror symmetry with respect to the time reversal. Clearly, it is due to two factors: 1$^{\circ}$- symmetry of the function $h_{o}(t)$ with respect to the time reversal $h_{o}(-t)=h_{o}(t)$, 2$^{\circ}$- the absence of barriers for nucleation of dislocation loops.

In real experiments, where due to the finite tension of dislocations nucleation barriers exist, this symmetry with respect to the time reversal is broken. This is obvious in Fig.\ref{fig:serial}b representing the spatio-temporal cross section extracted from a video of the experiment along the line CS defined in Fig.\ref{fig:serial}a-t4. When compared with the theoretical barrier-free scheme in Fig.\ref{fig:serial}c, the right part of Fig.\ref{fig:serial}b (expansion sequence) is similar to the theoretical barrier-free scheme because the collapse of dislocation loops is free. The left part of Fig.\ref{fig:serial}b (compression sequence) is very different from the theoretical scheme because the nucleation of new dislocation loops in the experiment is protected by energy barriers.

Let us analyse the spatio-temporal cross section in Fig.\ref{fig:serial}b at t=t1 where the cholesteric droplet is free of dislocations. This state of the cholesteric droplet can be qualified as \emph{multimetastable} because four dislocation with indices N=20, 19, 18, and 17 are missing in it so that the cholesteric helix in the center of the gap is supertwisted: the thickness in the center of the gap is $h_{o}=15.5p_{o}$ but it contains 17 $2\pi$ turns of the helix. The real pitch p of the helix equal to $p=(15.5/17)p_{o}$ is thus shorter than the equilibrium pitch $p_{o}$.

The elastic energy per unit volume of this supertwisted state at t=-1000s is given by:
\begin{equation}\label{eq:energy_disloc}
F=\frac{K_{22}}{2}\left(\frac{2\pi}{p}-\frac{2\pi}{p_{o}}\right)^{2}=\frac{2\pi^{2}K_{22}}{p_{o}^{2}}\left(\frac{p_{o}}{p}-1\right)^{2}
\end{equation}
The ratio $p_{o}/p$ called \emph{compression ratio CR} measures the supertwist distortion of the cholesteric helix.
\subsubsection{Nucleation of individual dislocations at the meniscus}
\label{sec:nucleation_meniscus}
During the subsequent compression of the gap, the supertwist strain increases and nucleation of several pairs of thin lines (equivalent to one thick line) occurs. These nucleation events occur at critical values of the compression ratio which can be estimated graphically in Fig.\ref{fig:serial}b as the ratio of lengths of the blue and red thick lines. Assuming that variation of the gap thickness in time is linear, the length of the blue line is proportional to the thicknesses $h_{o}$ at which dislocations with the index N would be in equilibrium ($F_{P-K}=0$). Similarly, the length of the red line is proportional to the gap thickness at which nucleation of the N$^{th}$ dislocation occurs. The critical compression ratios for nucleation of the dislocation pairs with indices N=15, 14 and 13 estimated by this method are respectively  CR=1.13, 1.2 and 1.18.

Let us stress that all these dislocations nucleated at the meniscus sweep the area of drop and come out from the drop on the other side of the meniscus.
\subsubsection{Serial nucleation of loops on dust particles}
\label{sec:nucleation_serial}
Next nucleation event occurring at $t=t_{2}$ (see Fig.\ref{fig:serial}b) is very different. Nucleation takes place on a dust particle located in the center of the gap and it is multiple: seven dislocations are nucleated in one burst. The critical compression ratio estimated graphically from Fig.\ref{fig:serial}b is CR=5.12 i.e. about five times higher than that of nucleation at the meniscus.

The subsequent nucleation event occurring at $t=t_{2}$ is also serial: four dislocations are nucleated in one burst (see Fig.\ref{fig:serial}b  and Fig.\ref{fig:serial}a-t3) at the compression ratio CR=4.86.
\begin{figure*}
\begin{center}
\includegraphics[width=4.8 in]{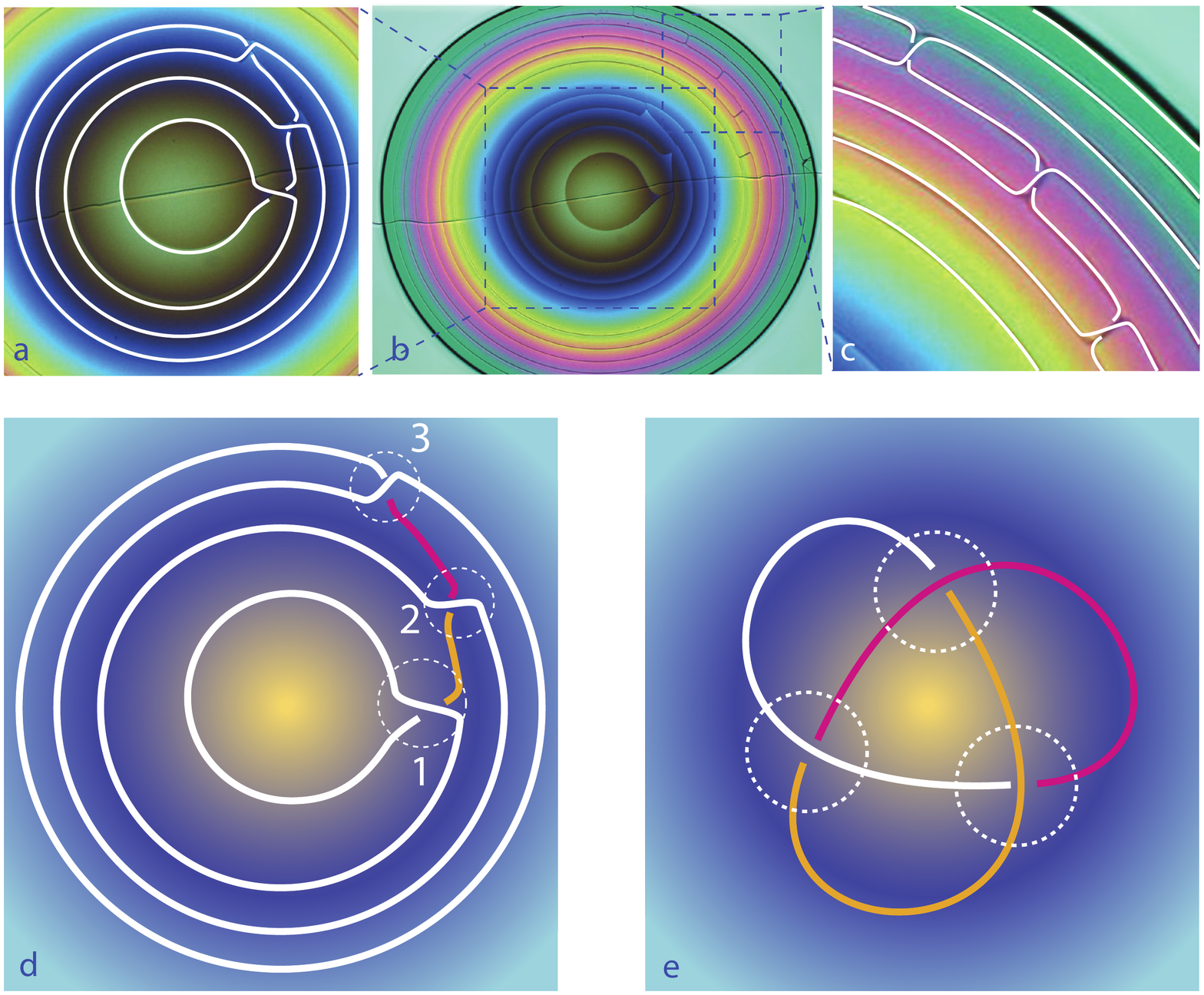}
\caption{Structure of the dislocation superloops generated by the serial nucleation events. a) Structure of crossings linking four loops into one superloop. b) Superloop generated by the serial nucleation of four loops at t=t3 in the experiment represented in Fig.\ref{fig:serial}. c) Structure of crossings linking seven loops into one superloop. d) Unsuccessful attempt to color the superloop from the picture a with three colors. e) Tricolorability of the trefoil knot.}
\label{fig:serial_superloop}
\end{center}
\end{figure*}
\section{Dislocation superloops}
\label{sec:knots}
\subsection{Dislocation loops linked into superloops}
\label{sec:superloops}
Dislocation loops generated in one series are not independent but linked by crossings into one closed line that we propose to call \emph{superloop}. This remarkable feature is illustrated in Fig.\ref{fig:serial_superloop} in which the picture b is the same as the one in Fig.\ref{fig:serial}a-t3. Let us consider first the four loops created in the event that occurred at t=t3. They are redrawn in Fig.\ref{fig:serial_superloop}a with the aim to unveil the structure of the three crossing linking adjacent loops. Clearly, the four loops are linked by the three crossing into one superloop.

Similarly, the drawing in Fig.\ref{fig:serial_superloop}c unveils the structure of crossings linking adjacent loops in the series of seven loops generated at t=t2 (see Fig.\ref{fig:serial}a-t2). Here, the seven loops are linked by six crossings (only three of them are visible in this picture) into one superloop too.
\subsection{Are superloops knotted ?}
\label{sec:unfolding_superloops}
Dislocation superloops are thus \emph{selfintersecting closed lines} exactly like the 2D projections of 3D knots are.

In the theory of knots, the number of crossings is the parameter used in classification of knots. The so-called trefoil knot shown in Fig.\ref{fig:serial_superloop}e is the unique knot with three crossings. One can ask therefore if the superloop with three crossings shown in Fig.\ref{fig:serial_superloop}a has topology of the trefoil knot. Several methods can be used to answer this question.
\subsubsection{Tricolorability}
\label{sec:unfolding_superloops}

One of them is to use the invariant called tricolorability. The trefoil knot in Fig.\ref{fig:serial_superloop}e is drawn using three colors in such a manner that colors of the three strands meeting at each crossing are different. The superloop in  Fig.\ref{fig:serial_superloop}a cannot be colored in this manner because only in the crossing labeled 2 the three strands have different colors while in the other two crossings, 1 and 3, the strands have only two colors.
\subsubsection{Unfolding of the superloop}
\label{sec:unfolding_superloops}
\begin{figure*}
\begin{center}
\includegraphics[width=5 in]{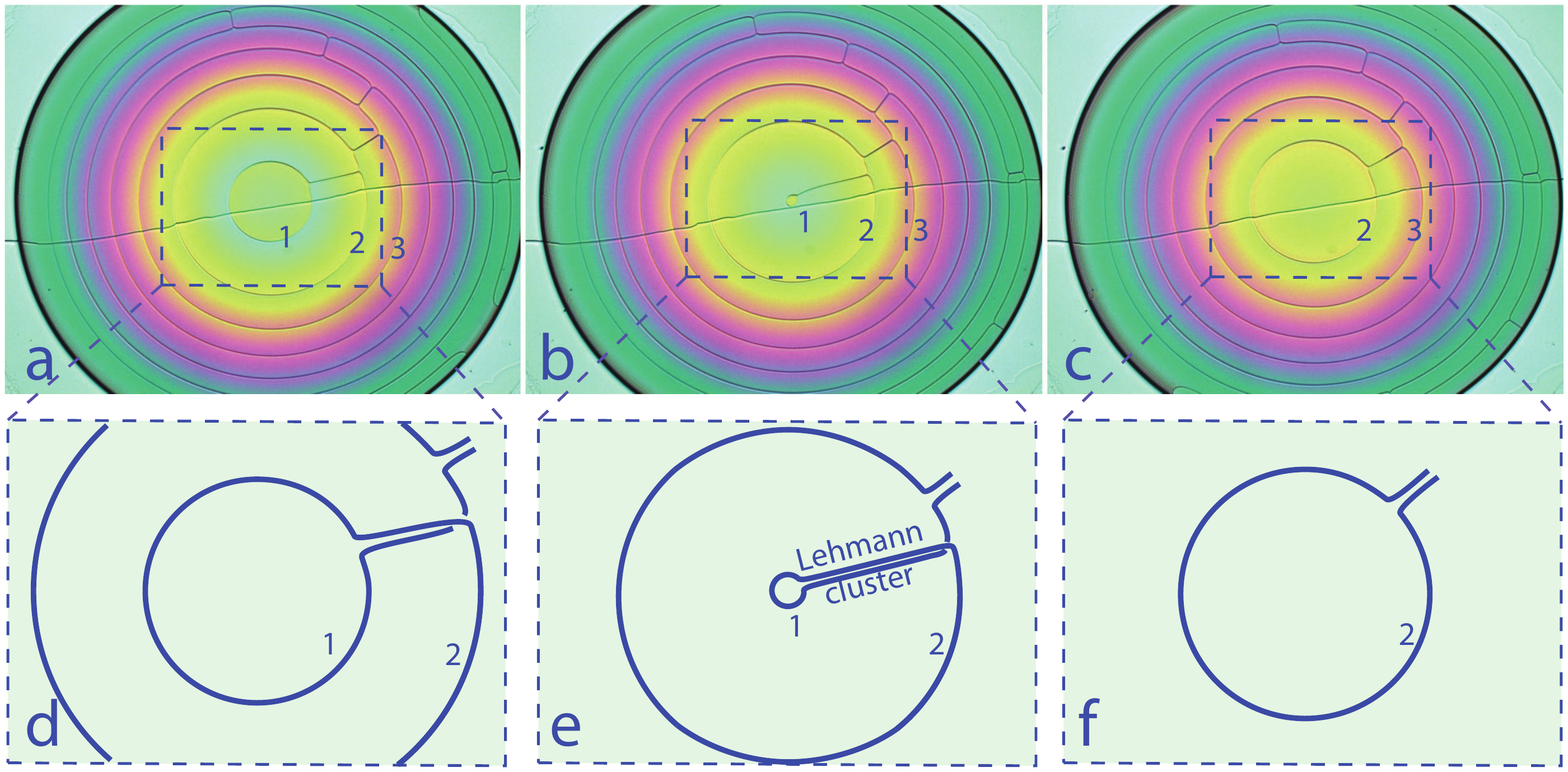}
\caption{Unfolding of a superloop during expansion of the cylinder/cylinder gap. a-c) Experiment: continuous suppression of the smallest loop of the folded superloop. d-f) Detailed representation of the unfolding process.  }
\label{fig:superloop_bis}
\end{center}
\end{figure*}

The second method is experimental and it consists in a continuous transformation of a superloop into a single loop without crossings, i.e. the unknot. This transformation is achieved by an expansion of the gap. An example of such a transformation applied to the superloop with six crossings is given in Fig.\ref{fig:superloop_bis}. The three pictures in Figs.\ref{fig:superloop_bis}a-c show that during the expansion of the gap, loops linked by crossings into the superloop, shrink. Let us follow the deformation of the smallest loop labeled 1. It is represented by three drawings in Figs.\ref{fig:superloop_bis}d, e and f. First, its radius shrinks. Subsequently, the residual hairpin-like segment, which is nothing else but the elementary Lehmann cluster \cite{Smalyukh_Lavrentovich_2002}, shortens to zero so that the loop is suppressed. \emph{N.B: In the theory of knots, the deformation leading to the elimination of one loop is called the Reidemeister move of type I}.

In the remaining configuration shown in Fig.\ref{fig:superloop_bis}f, the loop labeled 2 is linked  by one crossing with the loop 3 in the same manner as the loops 1 and 2 were linked in Fig.\ref{fig:superloop_bis}a so that the deformation described above can be repeated. By this means, superloops can be continuously unfolded into a single loop without crossings.
\begin{figure*}
\begin{center}
\includegraphics[width=5 in]{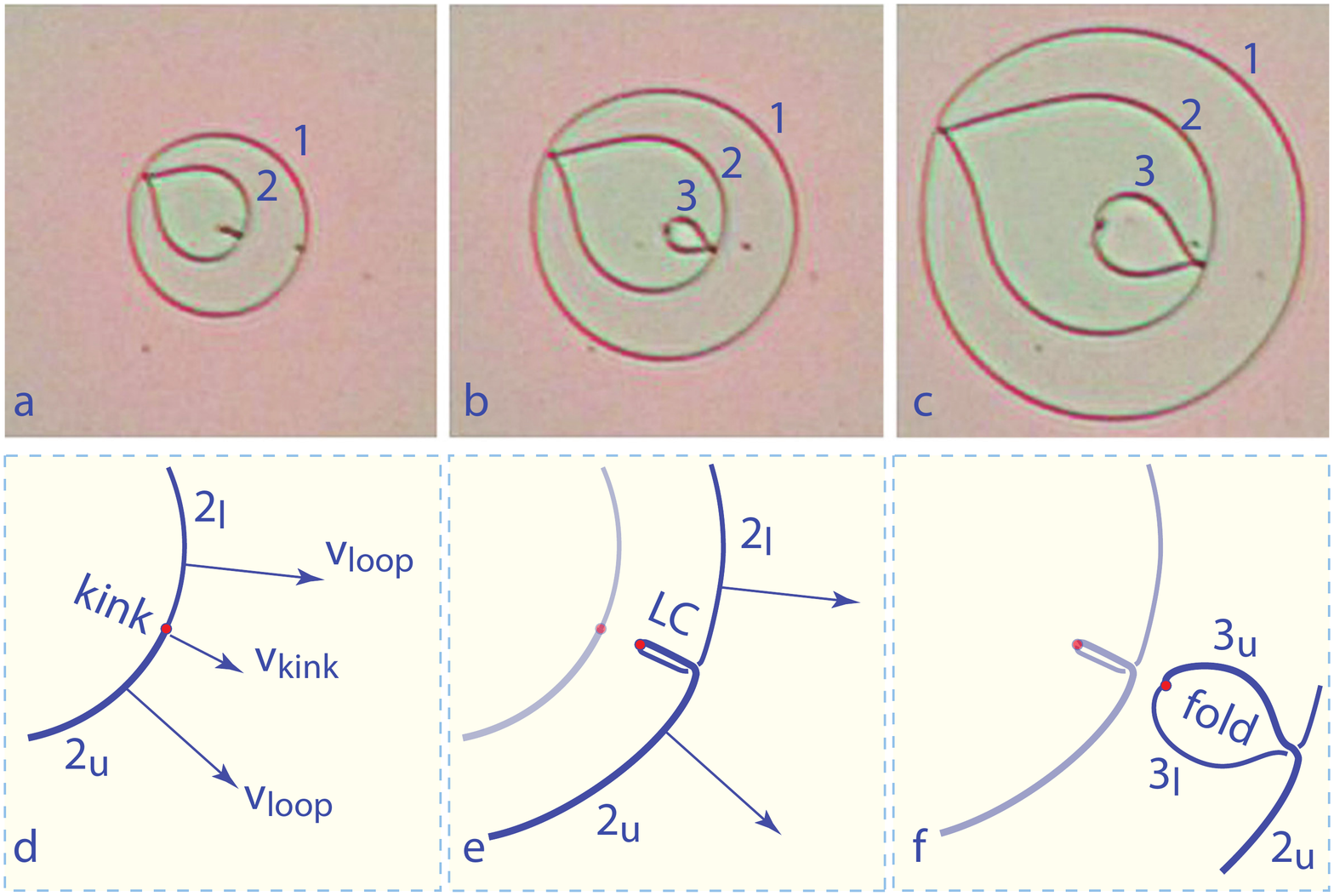}
\caption{Generation the third loop by folding of the second loop during a serial nucleation of a superloop. a-c) Experiment. d-f) Detailed representation of the folding process.    }
\label{fig:folding}
\end{center}
\end{figure*}
\subsubsection{Folding of loops into superloops}
\label{sec:folding_superloops}
The third method is experimental too and it refers to generation of superloops. It is illustrated in Fig.\ref{fig:folding} by the series of three pictures a, b and c showing generation of the third loop during a serial nucleation. The series of three drawings d e and f explains that the kink (possibly carrying a dust particle) plays the crucial role in the serial nucleation process. The necessity of the presence of kinks on dislocations in cholesterics has been explained in ref.\cite{Pieranski_Memorial}.

In the first drawing in Fig.\ref{fig:folding}d, the kink is marked with a red circle and the segments of the loop 2 adjacent to the kink are drawn with lines of different thicknesses because they are not located at the same level: the segment 2$_{l}$ is below the segment 2$_{u}$. Due to the compression of the wedge, the cholesteric helix is supertwisted and the Peach-Koehler force acting on the dislocation drives the expansion of the loop. The velocity of the expansion $v_{loop}$ is represented by arrows. The second drawing in Fig.\ref{fig:folding}e shows that as the kink introduces an additional dissipation, its velocity is lower than that of the loop, $v_{kink}<v_{loop}$, so that it stays more and more behind the loop during its expansion. The kink remains connected to the loop by the elementary Lehmann cluster (a metastable state), i.e. by a pair of associated dislocations, of growing length. The third drawing in Fig.\ref{fig:folding}f shows that upon a large enough compression ratio the Lehmann cluster splits into a the new loop labeled 3. \emph{N.B: In the theory of knots, this deformation leading to a continuous generation of an additional loop is called the Reidemeister move of type I}.

\subsubsection{Terminal compression ratio of serial nucleations}
\label{sec:folding_superloops}
Let the critical compression ratio for the splitting of the Lehmann cluster be $CR_{LC}$. It is known from experiments that we will report elsewhere that $CR_{LC}\approx 2.3$. As long as the compression ratio is larger than $CR_{LC}$ the serial nucleation continues.

From experiments reported in section \ref{sec:nucleation_serial} we know that the critical compression ratio for the beginning, at t=t2, of the serial nucleation of seven loops indexed N=13,...,7 is $CR_{serial}=5.12$. The thickness of the gap at which the serial nucleation starts is thus $h/p_{o}=5.12/13$. After nucleation of each loop the compression ratio is reduced to the value given by $CR_{N}=CR_{serial} (N-1)/13$. Inequality $CR_{n}>CR_{LC}$ is satisfied when $N>13(CR_{LC}/CR_{serial})+1$ i.e. when $N>6.8$. For this reason the serial nucleation that started at t=t2 stops after nucleation of the loop with the index N=7.

\section{Generation of folded dislocation superloops in hypotwisted cholesterics}
\label{sec:nucleation_hypotwisted}
\subsection{Theoretical considerations}
\label{sec:nucleation_hypotwisted}
The folded dislocation superloops discussed above were generated by a compressive strain of the cholesteric layer confined between the crossed cylindrical mica sheets. During the serial nucleation on dust particles the compressive stress was relaxed due to the reduction of the initial number N of cholesteric pitches by increments of $\Delta N=-1$.

Theoretically, a similar mechanism of the serial nucleation on dust particles could be also driven by a large enough dilation of the cholesteric layer. However, as we know already from section \ref{sec:with_barriers}, in the crossed cylinders geometry, the dilation stress is relaxed by the reduction of the size followed by the collapse of dislocation loops already present in the gap.

Nevertheless, the number of dislocation loops inside the drop depends on its size. A small drop contains very few dislocation loops so that after the collapse of all of them under the dilation, nucleation of new loops occurs as we will point out below.
\subsection{Experiment: nucleation followed by folding }
\label{sec:nucleation_hypotwisted}
\begin{figure*}
\begin{center}
\includegraphics[width=5.5 in]{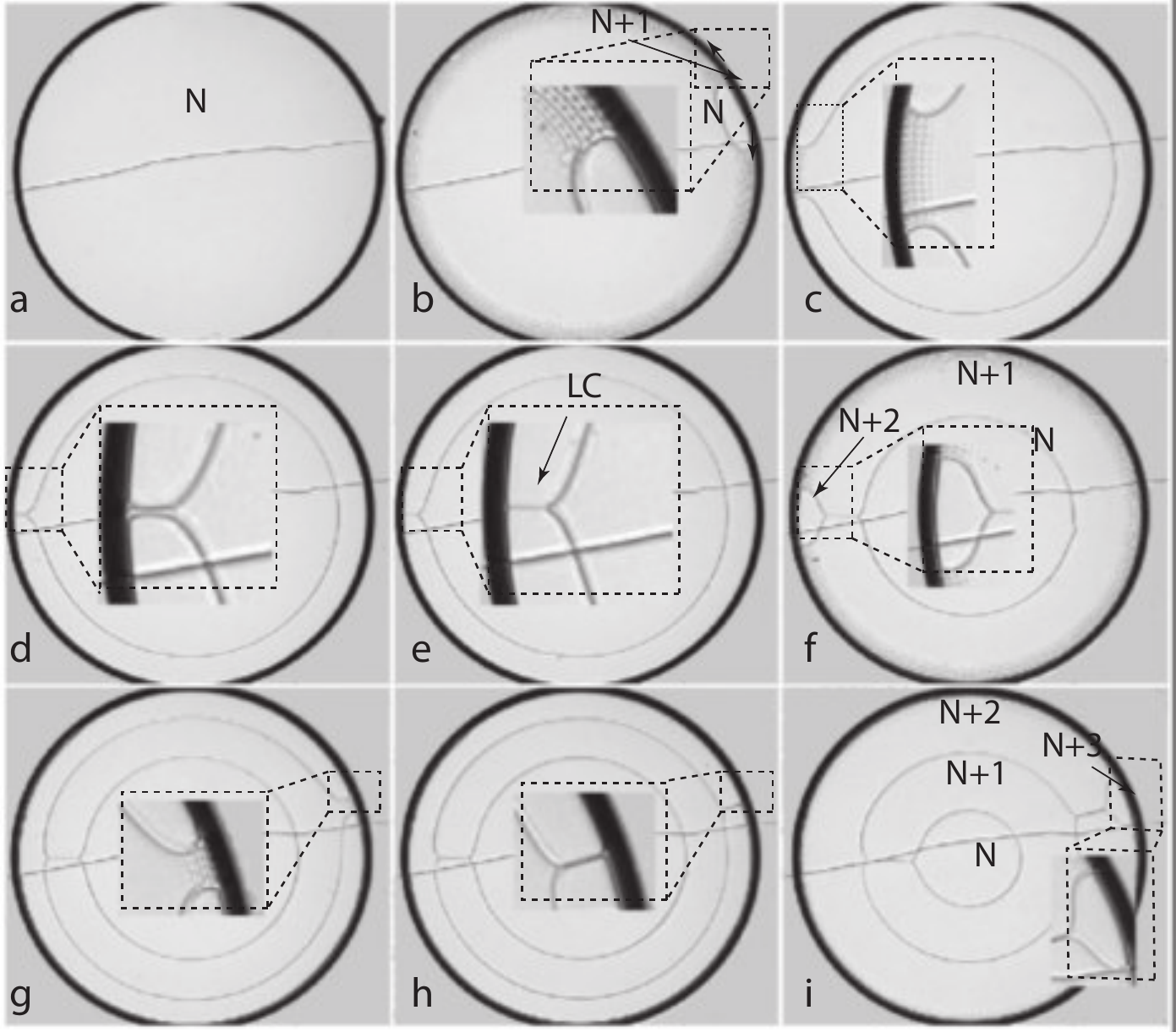}
\caption{Iterative generation of a folded superloop driven by dilation of the cholesteric layer. a) Drop free of dislocation. It contains N cholesteric pitches. b) Nucleation of a dislocation loop at the meniscus of the cholesteric droplet. Inside the loop the number of cholesteric pitches is N+1. b-c) Angular extension of the narrow and curved dislocation loop. d) Lateral segments of the dislocation loop came to contact. d) Transformation of the dislocations pair into the Lehmann cluster (LC). f) Generation of the field N+2 due to the splitting of the Lehman cluster. g-i) Generation of the field N+3.}
\label{fig:dilation_superloop}
\end{center}
\end{figure*}

Let us consider the series of nine pictures in Fig.\ref{fig:dilation_superloop} illustrating an experiment with a very small drop.

In the first picture (a), the droplet is free of dislocations and contains N cholesteric pitches. The next picture (b) shows that upon a dilation: 1$^{\circ}$ - the buckling instability took place at the edge of the droplet where the dilation ratio $(h_{o}+r^2/2R)/(Np_{o})$ is the largest and 2$^{\circ}$- nucleation of a dislocation loop occurred at the droplet edge.

The elongated curved shape of the loop results from the balance of the Laplace and Peach-Koehler forces:
\begin{equation}\label{eq:balance_L_PK}
F_{PK}-T\kappa=0
\end{equation}
in which $T$ is the tension of the dislocation and $\kappa$ its local curvature.

When the Peach-Koehler force overcomes the Laplace force, the angular extension of the loop increases in time. In the picture (c), it is close to $2\pi$, then in (d) the lateral segments of the dislocation loop came into contact. The next picture (e) shows that the pair of dislocation segments in contact is transformed into the Lehmann cluster (LC). Upon a further increase of the dilation ratio, the buckling instability occurs again at the edge of the droplet in the field N-1 and the Lehman cluster splits into the new loop labeled ``N+2".

The angular extension of the field labeled N+2 increases in time until its lateral segments come to contact in the picture g. This time, the transformation of the dislocations pair into the Lehmann cluster (LC) does not occur (picture h). Upon the next dilation step, the lateral segments of the field N+2 cross each other and the new field labeled N+3 is created (picture i). In the same manner, fields with indices N+4, N+5, ... can be generated by consecutive dilation steps. Simultaneously fields labeled N, N+1, ... will collapse in the drop center.

Let us emphasize that an evidence of this iterative mechanism of generation of new fields at the meniscus was present already in the spatio-temporal cross section in Fig.\ref{fig:serial}. Indeed, the shape of the two fields adjacent to the meniscus and indicated with arrows in the picture (a-t4) is the same as that of fields labeled ``N+2" and ``N+3" in Fig.\ref{fig:dilation_superloop}f and i.

\section{Confinement of knotted dislocations in the crossed cylinders geometry }
\label{sec:knots}
\begin{figure*}
\begin{center}
\includegraphics[width=5.0 in]{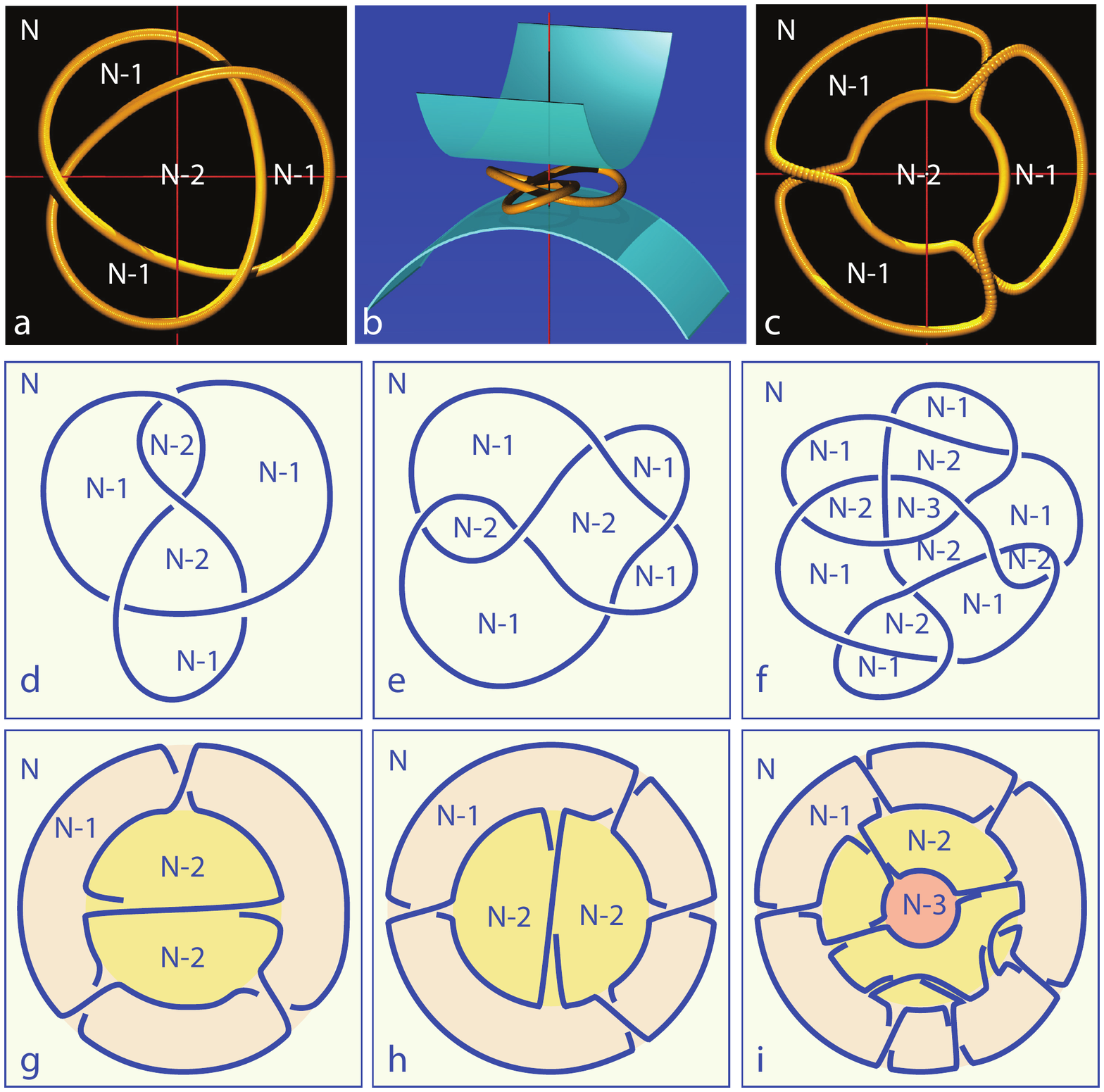}
\caption{Gedanken experiment with knotted superloops. a) Arbitrary shape of the trefoil knot. b) Confinement of the trefoil knot in the cylinder/cylinder gap. c) Shape of the trefoil $3_{1}$ knot confined in the cylinder/cylinder gap. d-f) Arbitrary shapes of the $4_{1}$, $5_{1}$ and Conway knots. g-i) Shapes of the $4_{1}$, $5_{1}$ and Conway knots confined in the cylinder/cylinder gap. Dislocations separate fields with different numbers N of cholesteric pitches. The number N decreases by increments $\Delta$N=-1 on a path from outside to the center of knots.     }
\label{fig:trefoil_gap}
\end{center}
\end{figure*}
From the formal point of view, dislocation superloops satisfy the definition of knots: closed selfintersecting lines with some number of crossings. However, as we have pointed out above, superloops generated by serial nucleation events are topological equivalent to the unknot - a single closed loop without crossings.

Let us suppose that by some methods which remain to be found it would be possible to generate more complex superloops having topology of knots. One could ask then what shapes would take such knotted dislocation superloops if they were confined between crossed cylinders.

This question is similar to the one concerning shapes of fluctuating knotted polymers formulated by P.G. de Gennes \cite{PGG}. He postulated that mean shapes of knotted polymers are constrained by maximisation of the entropy. Numerous works were devoted to the search for these so-called ideal knots (see for example in ref. \cite{Piotr_Pieranski}).

In the field of liquid crystals, a considerable amount of work was done on knots made of disclinations in chiral nematic colloids containing spherical \cite{Tkalec} or knotted inclusions \cite{Smalyukh_knots}. However, as far as we know, the issue of shapes of knotted dislocations in cholesteric layers confined in a cylinder/cylinder gap has not been raised so far elsewhere.

The simplest way to tackle it consists, first of all, in attribution of the number of cholesteric pitches to all fields delimited by dislocations. Let us start for example with the trefoil knot shown in Fig.\ref{fig:trefoil_gap}a and suppose that in the field outside of the knot the number of cholesteric pitches confined in the cylinder/cylinder gap is N. On the path from outside to the center of the knot (where the number of pitches is minimal), upon each crossing of the dislocation line the number of cholesteric pitches decreases by increments $\Delta$N=-1 (see Fig.\ref{fig:trefoil_gap}a). By this method applied to the trefoil knot one finds three field with the same index N-1 and one field, in the center of the knot, with the index N-2.

Due to the action of the Peach-Koehler force, dislocations separating adjacent fields would tend to follow loops (circles) with radii expressed by equation \ref{eq:rdisloc}. The crossings connecting adjacent dislocation loops would consist of pairs of dislocations with the total Burgers vector equal to zero. Dislocations in these pairs can be either dissociated or associated into Lehmann clusters.

The same method applied to the $4_{1}$, $5_{1}$ and Conway knots represented in Figs.\ref{fig:trefoil_gap}d-f, leads to shapes given in Figs.\ref{fig:trefoil_gap}g-i.
\section*{Acknowledgements}
The new experimental setup tailored for production of cylinder/cylinder mica wedges was built by V. Klein, J. Sanchez and S. Saranga. We benefitted from discussions with P. Oswald, O. Lavrentovich and P. Palffy-Muhoray as well from the help of I. Settouraman, M. Bottineau, J. Vieira, I. Nimaga, C. Goldmann and J. Saen.

\section*{Funding}
This work was co-financed by FEDER, European funds, through the COMPETE 2020 POCI and PORL, National Funds through FCT – Portuguese Foundation for Science and Technology and POR Lisboa2020 under projects PIDDAC (POCI-01-0145-FEDER-007688, Reference UIDB/50025/2020-2023).

\end{document}